\documentclass[a4paper,11pt]{article}
\usepackage{jinstpub} 
\usepackage{lineno}

\proceeding{Innovative Particle and Radiation Detectors 2023 (IPRD23)\\
 25-29 September 2023\\
 Siena, Italy}

\title{Pixel detector hybridisation and integration 
with anisotropic conductive adhesives}

\author[a,1]{Alexander Volker,\note{Corresponding author.}}
\author[a]{Janis Viktor Schmidt,}
\author[a]{Dominik Dannheim,}
\author[a]{Peter Svihra,}
\author[b]{Mateus Vicente Barreto Pinto,}
\author[a]{Rui de Oliveira,}
\author[a]{Justus Braach,}
\author[a]{Xiao Yang,}
\author[c]{Marie Ruat,}
\author[c,g]{Débora Magalhaes,}
\author[d]{Matteo Centis Vignali,}
\author[e]{Giovanni Calderini,}
\author[f]{Helge Kristiansen,}

\affiliation[a]{CERN, Meyrin, Switzerland}
\affiliation[b]{Universite de Geneve, Geneva, Switzerland}
\affiliation[c]{ESRF, Grenoble, France}
\affiliation[d]{FBK, Trento, Italy}
\affiliation[e]{LPNHE, Paris, France}
\affiliation[f]{Conpart AS, Skjetten, Norway}
\affiliation[g]{DESY, Hamburg, Germany}

\emailAdd{alexander.volker@cern.ch}

\abstract{A reliable and cost-effective interconnect technology is required for the development of hybrid pixel 
detectors. The interconnect technology needs to be adapted for the pitch and die sizes of the 
respective applications. This contribution presents recent results of a newly 
developed in-house single-die interconnection process based on Anisotropic Conductive Adhesives 
(ACA). The ACA interconnect technology replaces solder bumps with conductive micro-particles 
embedded in an epoxy layer applied as either film or paste. The electro-mechanical connection 
between the sensor and ASIC is achieved via thermocompression of the ACA using a flip-chip device 
bonder. The ACA technology can also be used for ASIC-PCB/FPC integration, replacing wire bonding or 
large-pitch solder bumping techniques. A specific pixel-pad topology is required to enable the 
connection via micro-particles and create cavities into which excess epoxy can flow. This pixel-pad 
topology is achieved with an in-house Electroless Nickel Immersion Gold (ENIG) process. The ENIG and ACA processes are qualified with a variety of different ASICs, 
sensors, and dedicated interconnect test structures, with pad diameters ranging from 12 $\mu$m to 140 
$\mu$m and pitches between 20 $\mu$m and 1.3 mm. The produced assemblies are characterized electrically, 
with radioactive-source exposures, and in tests with high-momentum particle beams. This 
contribution introduces the developed interconnect and plating processes and showcases different 
hybrid assemblies produced and tested with the above-mentioned methods. A focus is placed on 
recent optimization of the plating and interconnect processes, resulting in an improved plating 
uniformity and interconnect yield.}


\begin{document}
\maketitle
\flushbottom

\section{Introduction}
\label{sec:intro}

Interconnect technologies from fine pitch hybridisation to large pitch module packaging are important for a wide range of applications in the field of high-energy physics and other silicon pixel-detector applications. In many cases state of the art is solder bump bonding in which the electrical connection between metallized pads is created by solder balls under reflow conditions. This requires many steps of preparation including a process called Under Bump Metallization (\textbf{UBM}) which are done on a wafer level prior to dicing. For small-scale applications and during the ASIC and sensor development phase, interconnect technologies must also be suitable for the assembly of single-dies typically available from Multi-Project-Wafer submissions. For these multi project wafers, wafer level processes are either not possible or too costly for the scale of the project. Within the CERN EP R\&D\cite{EPRND} program and the AIDAinnova collaboration\cite{AIDAinnova}, innovative and scalable hybridisation concepts are under development for pixel-detector applications in future colliders. An innovative approach for in-house single die bonding is discussed in this contribution. The electrical interconnection is created using Anisotropic Conductive Adhesives (\textbf{ACA}) in the form of conductive particles embedded in an epoxy paste or film and by thermo-compressing the adhesive. Commercial use of these types of adhesives can be found in display panel integration\cite{Panel} and also increasingly in chip-to-film\cite{COF} and chip-to-flex bonding\cite{Svihra, Schmidt}. The required surface properties of the devices bonded with ACA are achieved with an in-house single-die plating process.

\section{Electroless Plating}
\label{sec:ENIG}

\begin{figure}[htbp]
\centering
\includegraphics[width=.4\textwidth]{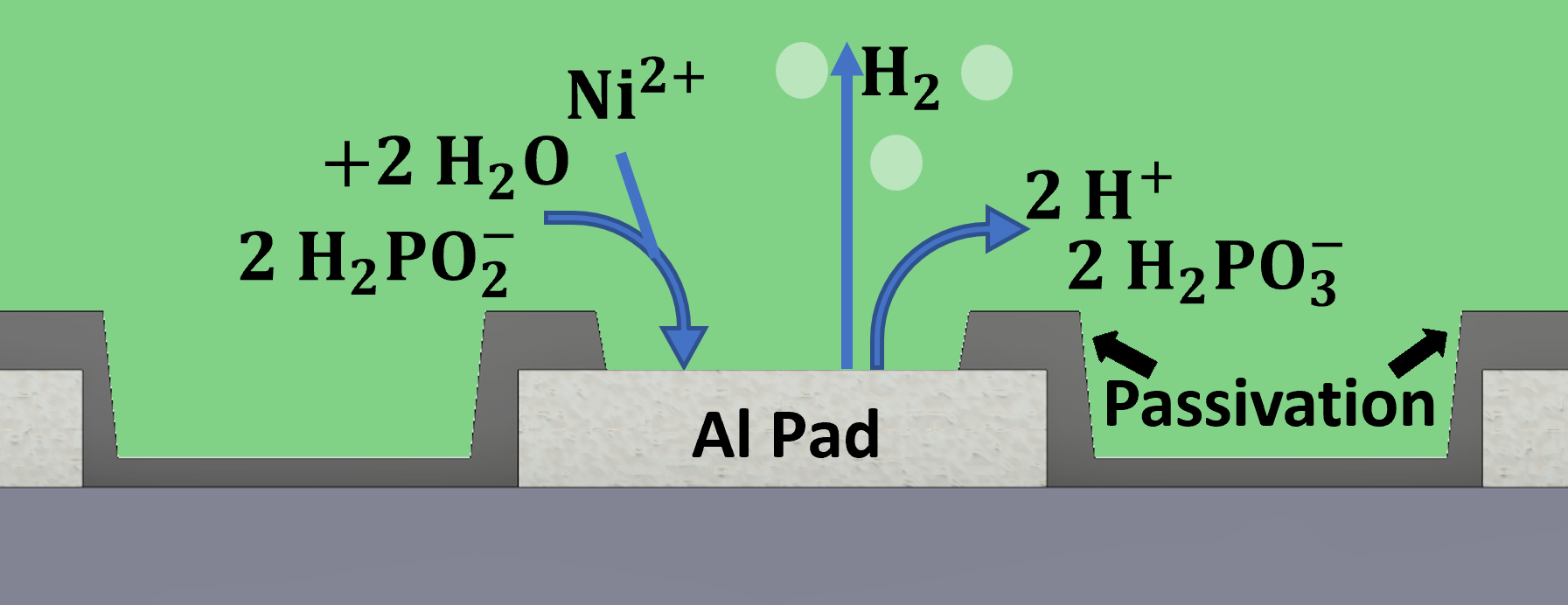}
\caption{The graphic shows the simplified chemical process in the Electroless Plating bath when plating nickel on aluminium pads that are partially covered by a passivation layer.
\label{fig:ENIG}}
\end{figure}

For the topology requirement needed to bond with the ACA, Electroless Nickel Immersion Gold (\textbf{ENIG}) plating is utilized, a method typically used in PCB production\cite{PCB}. It is a chemical plating process that grows nickel on metallized pads without the need for an external electrical current, making it thereby useable for technologies operating on semiconductor electronics. In the case of ASICs and sensors, the pad metallization widely used is 
an aluminium alloy which first requires a pre-treatment of the pads in order for the ENIG plating to create a catalytic surface on which the reaction can start reliably. The first step after mounting the samples on a holder is plasma cleaning them. This removes part of the thin oxide layer and increases the wettability for further chemical processing. In the next step the oxide layer is chemically removed and more roughened surface for plating to adhere to is created. The next step is generally referred to as zincation. In this step a thin seed layer of Zinc is chemically deposited on the aluminium surface. The cycling of this step twice or thrice can increase the quality of the zinc seed layer. The sample can now be submerged in the Electroless Nickel bath containing hypophosphite and Ni-Ions to start the plating process as described by the chemical equation \ref{eq:Ni} and shown in Fig. \ref{fig:ENIG}. A deposition of metallized nickel is observed. Due to the auto catalytic nature of the process the nickel will grow radially outwards from all points of deposition, thereby creating a mushroom-like shape with increasing plating height.

\begin{equation}
\label{eq:Ni}
\begin{aligned}
\text{H}_2\text{PO}_2^- + \text{Ni}^{2+} + 2\text{H}_2\text{O} \rightarrow \text{H}_2\text{PO}_3^- + \text{H}_2 + 2\text{H}^+ + \text{Ni}^0
\end{aligned}
\end{equation}

\begin{figure}[htbp]
\centering
\includegraphics[width=0.5\textwidth]{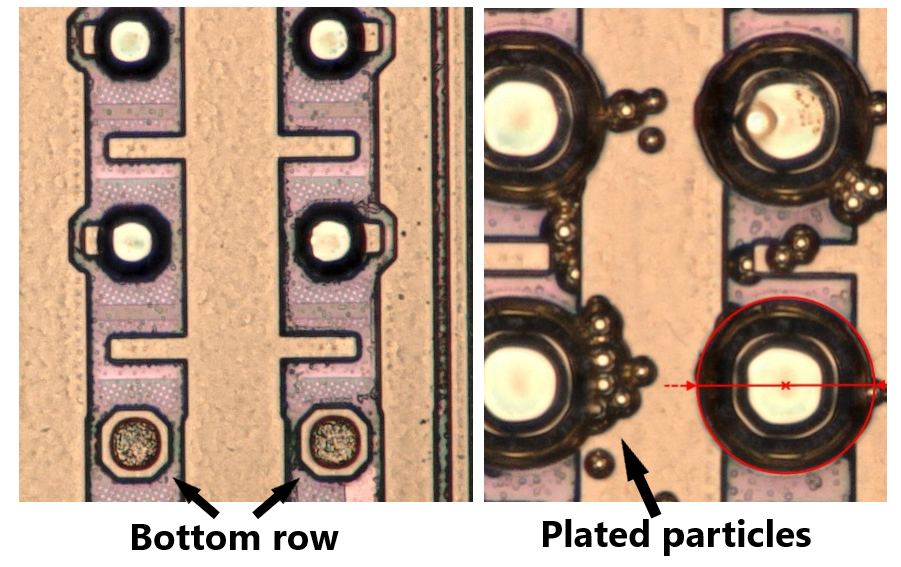}
\caption{Electroless Nickel defects on small pitch (55 $\mu$m)  devices. The left picture shows a electrically connected row of pads in the bottom of the pixel matrix. This row shows frequently different plating behaviour compared to the rest of the pixel matrix. The right picture shows plate out happening between pads. Small dots that increase in number and grow leading to shorts. 
\label{fig:Plating}}
\end{figure}

The plating height is controlled by the plating time while defects in the plating growth process limit the height at which uniform growth on all pads can be achieved. Defects observed are single pads not plated or skipped, pads near edges of the chip un-plated or considerably lower plating height. Observations show a plate out effect showing nucleation of nickel plating on surfaces where plating is undesired (see Fig. \ref{fig:Plating} right). This occurs more frequently when the 
plating gets high enough for the space between pads to shrink considerably. It starts a chain reaction creating increasingly more nucleation spots that each grow with further plating time. A lead-based stabilizer is used as catalyst poison in order to inhibit the plate out. Another challenge to achieve an even plating topology are pads with different metal structures and shorts. These pads can show different plating speeds compared to the rest of the pixel matrix leading to uneven topologies (see Fig. \ref{fig:Plating} left). \par

After the Electroless Nickel process an Immersion Gold process is used to create a thin gold layer on the previously plated nickel in order to protect them from corrosive conditions. The gold atoms replace the nickel atoms at the surface, slowly growing the gold layer. The chemical equation for this process is shown in \ref{eq:Au}.

\begin{equation}
\label{eq:Au}
\begin{aligned}
\text{Ni} + 2\text{Au}^+ \rightarrow \text{Ni}^{2+} + 2\text{Au}
\end{aligned}
\end{equation}

\begin{figure}[htbp]
\centering
\includegraphics[width=0.6\textwidth]{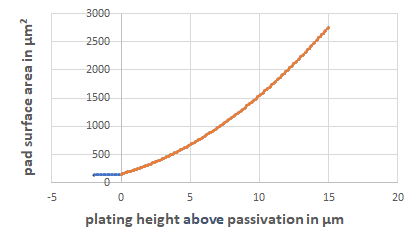}
\caption{Graph showing the calculated increase of Ni-pad surface with increased plating height assuming uniform radial growth. The initial Al-pad is 14$\mu$m in diameter and 2$\mu$m in height below the passivation layer.
\label{fig:Stirring}}
\end{figure}

Agitation of the Electroless Nickel bath is an important parameter to control as it strongly affects the 
diffusion zone in which the chemical processes of the plating take place\cite{Stirring}. A stirring hotplate setup is used where the stirring speed $\omega$ 
is adjusted according to the surface of the growing nickel bumps $\omega \sim h(h + 1)$ with $h$ for the height of the plating above the passivation. 
Due to the mentioned mushroom-like shape the surface area will grow quickly for small pad devices as shown in Fig. \ref{fig:Stirring}. For devices with pad radii larger than 50 $\mu$m, the increase in surface area gained during plating becomes progressively less significant and is thus of diminishing relevance in the process.  All measures of plating height are stated with reference to the passivation covering the edges of the pads. Adjusting the stirring results in more uniform topologies with less skipping as reported by Siau et al.\cite{Stirring2} and a slower onset of plate out. The agitation also reduces the formation of bubbles on the active surface.

\section{Bonding with Anisotropic Conductive Adhesives}
\label{sec:ACA}

\subsection{ACA materials}

While the Anisotropic Conductive Films (\textbf{ACF}) are commercial 
products with detailed bonding parameters (see Table \ref{tab:Materials} left), adjustments based on material and pixel pad layout still have to be done. This can be the case for large chips where it is required to squeeze out as much of the epoxy as possible prior to curing the epoxy. For the 
Anisotropic Conductive Paste (\textbf{ACP}), commercial epoxies with different conductive particles embedded are available for use based on 
material characteristics such as viscosity and radiation hardness needed for the device and bonding process. Mixing the conductive particles into 
the paste is also done in-house. In use are a range of particles in sizes and coating materials as 
shown in Table \ref{tab:Materials} right.

\begin{table}[htbp]
\centering
\includegraphics[width=0.4\textwidth]{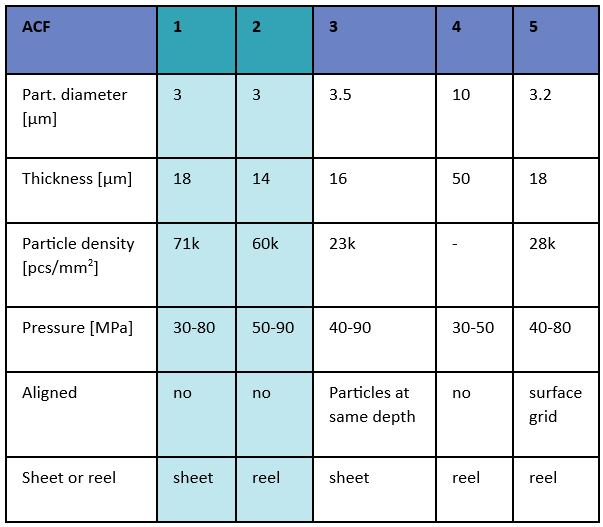}
\qquad
\includegraphics[width=0.4\textwidth]{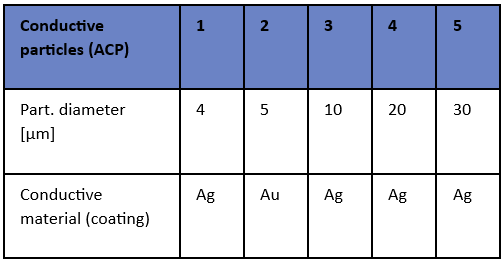}
\caption{(left) Table of commercially available ACF with highlighted number 1 and 2 due to their most common usage. (right) Table of conductive particles available for in-house ACP mixing. The numbering of the ACA products is for overview only and does not correspond to product names or manufacturers.
\label{tab:Materials}}
\end{table}

\subsection{Flip-chip bonding}

\begin{figure}[htbp]
\centering
\includegraphics[width=.7\textwidth]{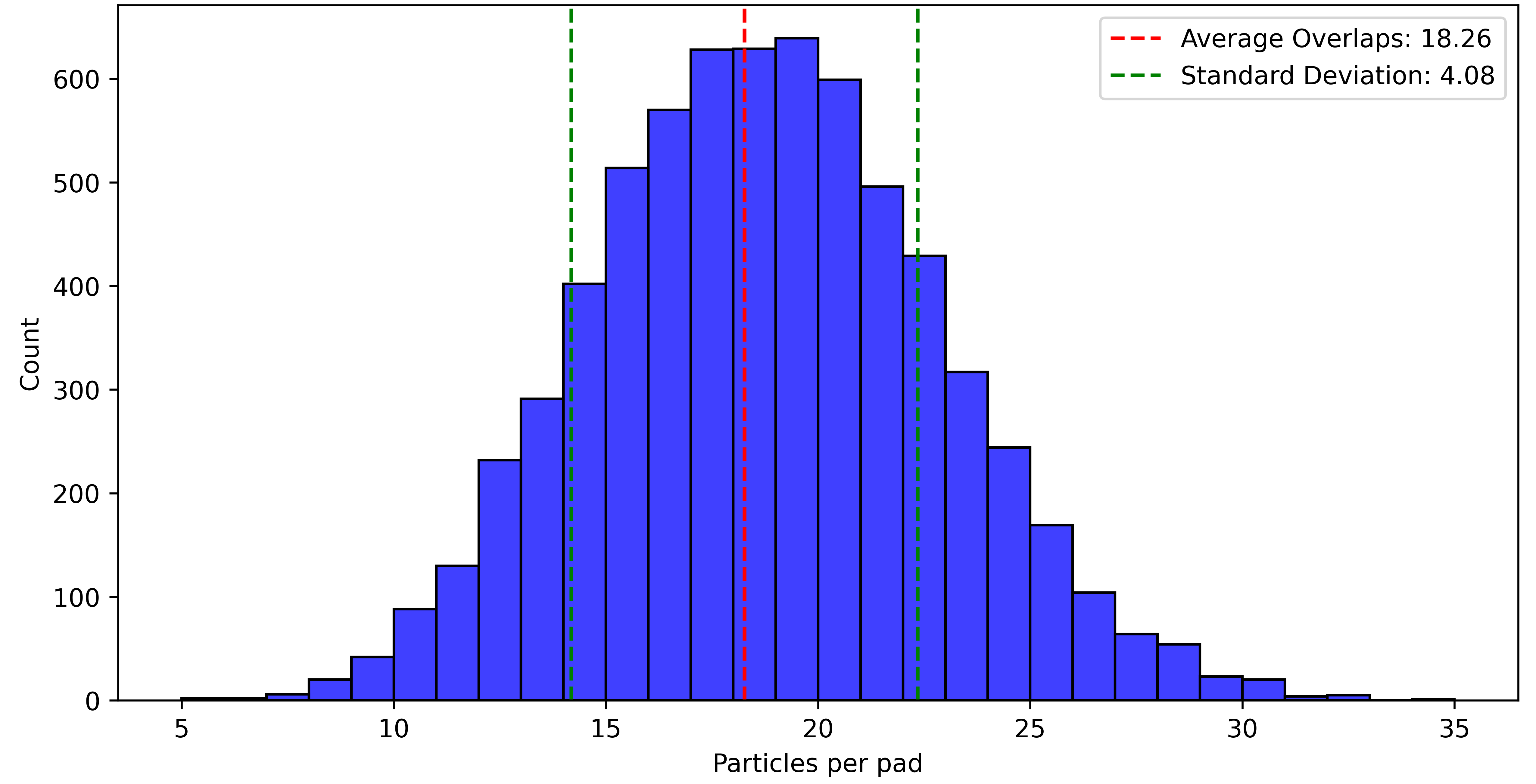}
\caption{Histogram of a simulation result showing the distribution of particles pinned per pad. The pads are 50 $\mu$m in radius while the particles are 10 $\mu$m in diameter. In this realisation a particle volume fraction of 0.10 is assumed. The total pad count is 6724.
\label{fig:Histogram}}
\end{figure}

ACF is applied to one side of the parts to be connected in a lamination process defined by the 
manufacturer, while ACP is dispensed by a syringe dispensing system.
The dispensing force and the speed of the syringe tip while dispensing is set to control the adhesive quantity. The quantity of adhesive dispensed is being reduced without creating uncovered pads due to skipping. For the ACF the 
quantity of adhesive is controlled by the thickness of the 
film. A too thick film of ACF will lead to an excess of high-viscosity 
adhesive resulting in areas of unconnected pixels. For most applications, in particular fine pitch ones, thinner films are therefore preferred. Newest developments in the ACF industry show film 
thicknesses down to 10 $\mu$m with embedded metal coated polymer particles of 3 $\mu$m diameter. Larger particles are required when connecting uneven structures as the particle itself can be 
deformed to less than half its diameter, therefore evening out the topological differences. Smaller particles in the 3-4 $\mu$m diameter range however are still in favor for most of fine pitch applications, as reliable 
connections require a high particle density that can not be realised efficiently with larger particles. For the commercially available ACF like ACF3 or ACF5, simulations of the particle distribution show pads of devices with fine pitch not having a high enough particle yield. An example of a simulation of an ACP particle distribution for a larger pitch device is shown in Fig. \ref{fig:Histogram}. The particle placement is picked randomly and then in case of overlap pushed away radially. This is done to emulate the conditions of particles that overlap in 2D top-down view but not 3D view to behave as if thermo-compressed. 

\par
\begin{figure}[htbp]
\centering
\includegraphics[width=.6\textwidth]{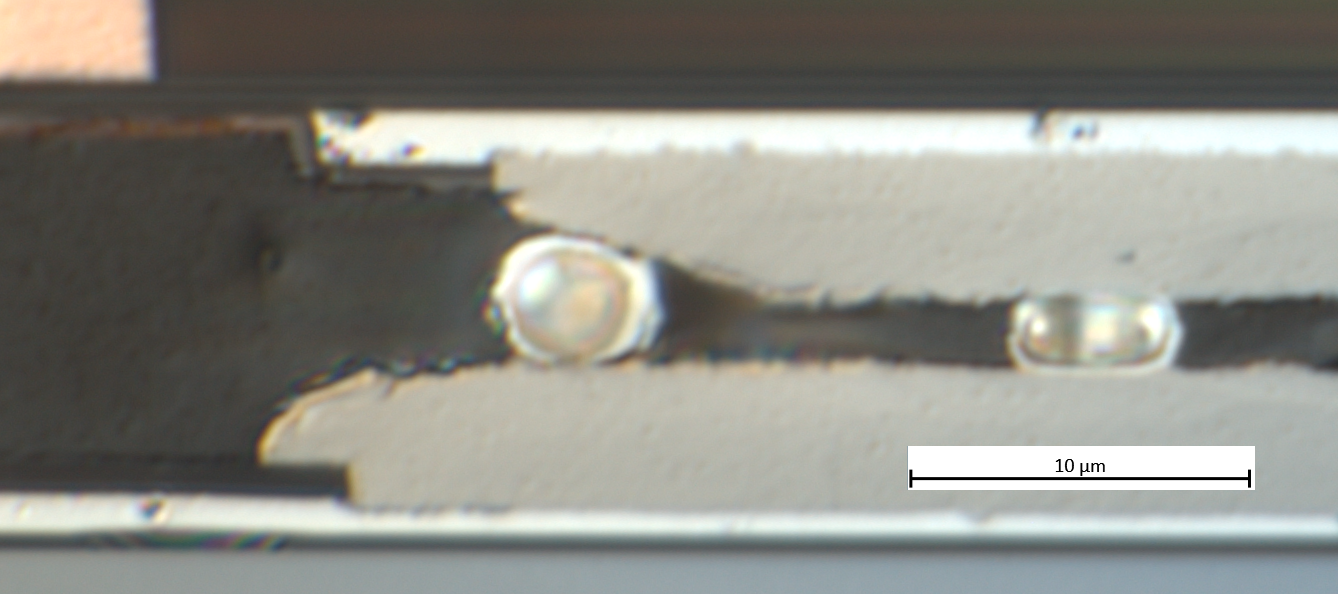}
\caption{Microscopy cross-section picture of an ACP bonded ALTIROC2 ASIC (top) and an LGAD sensor (bottom). Two conductive particles are pinned between the ENIG plated pads, therefore electrically connecting the two pads. The right particle is pinned between two parallel planes while the left one is pinned between two planes at an angle of about 20°.\label{fig:20°}}
\end{figure}

A cross-section analysis shows particles being pinned between nonparallel pad surfaces at an angle of 20° (see Fig. \ref{fig:20°}). This allows for particles being pinned beyond the nominal pad size as the ENIG plating creates round edges around the pad.

For the flip-chip process, a device bonder that can apply up to 100 kg of force and 
temperatures up to 400°C is used. The temperature used is dependent on the curing temperature of the 
epoxy used as adhesive for the ACA. For ACF1 and ACF2 150°C is used as curing temperature while bonding. The bonding pressure used can be as high as 90 MPa for the pad connection area with some devices requiring a lower pressure on the sensor material. For ACP the temperature and bonding pressure applied is generally lower and highly dependent on the materials used.   
The bonding force is applied till after the cooling of the sample back to room temperature.

\section{Results}
\label{sec:Results}

\subsection{ENIG plating}

Consistent plating results for large pitch (>1000 $\mu$m) substrates with large pads (>50 $\mu$m in diameter) have been achieved, showing uniform plating heights as desired for the applications. Plating heights for those substrates up to 20 $\mu$m have been achieved. Other types of substrates show defects limiting the plating to certain ranges depending on the specific layout. Small pitch (<50 $\mu$m) pixel matrices tend to plate out once growing pads decrease the gap between the pads. For small pads (<20 $\mu$m diameter) the plating is perturbed and therefore strongly influenced by bath parameters such as temperature, agitation, and concentration of bath constituents. This results in large-pad substrates with small pixel pitches showing plating resilient to perturbations, but early onset of plate out. The plating height for such substrates is therefore limited by the distance between the pixel pad edges. Substrates with large pitches and small pads are dominated by perturbations. Substrates with small pitches and small pads pose the biggest challenge for the current ENIG plating process. ENIG plating on chips with a pixel pitch of 50 $\mu$m and 7 $\mu$m pad radius has been achieved up to a plating height of 8 $\mu$m by carefully controlling stabiliser concentrations and stirring speeds while processing.
Table \ref{tab:heightdevice} displays the achievable plating heights on devices with varying pad sizes and pixel pitches.

\begin{table}[h]
\centering
\includegraphics[width=.6\textwidth]{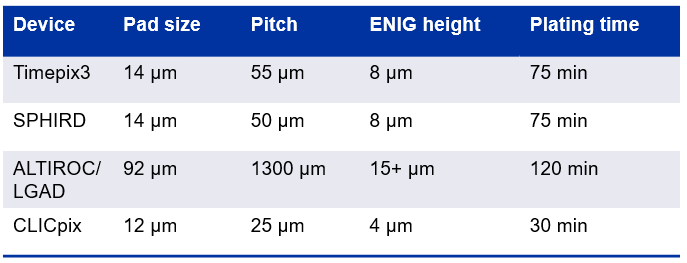}
\caption{Results of achievable plating heights for selected devices. The plating height on the ALTIROC/LGAD device is limited by the chosen plating time and not by the occurrences of any defects.  
\label{tab:heightdevice}}
\end{table}

\subsection{Test-Structure bonding}

\begin{figure}[h]
\centering
\includegraphics[width=.9\textwidth]{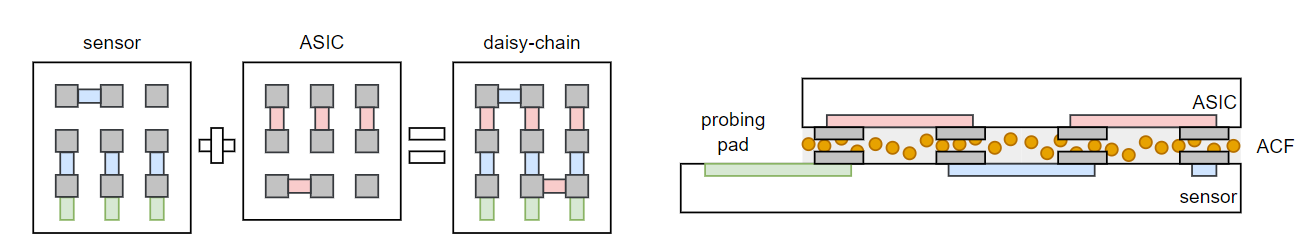}
\caption{Schematic of a daisy chain device for testing of ACA bonded interconnections.
\label{fig:chainexample}}
\end{figure}

\begin{figure}[htbp]
\centering
\includegraphics[width=1\textwidth]{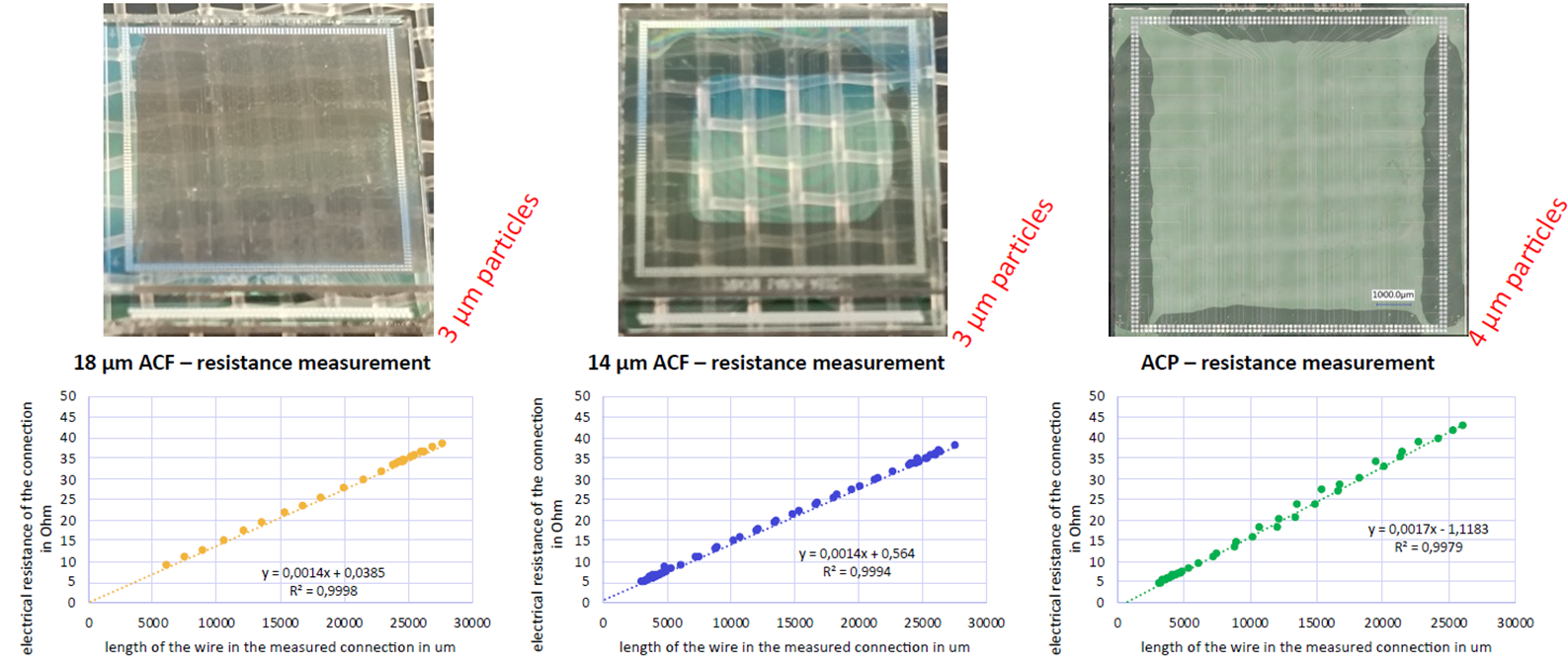}
\caption{Pictures of the three different daisy chain devices and their corresponding electrical resistance measurements. On the left is the ACF1 applied as sheet while the ACF used in the device in the middle is made with ACF2 applied in four separate stripes. For the assembly shown on the right the ACP was dispensed in four lines to get a good coverage without risking air entrapment.
\label{fig:chainperipheral}}
\end{figure}

For a systematic characterisation and optimisation of the plating and interconnect process, test devices with a daisy chain connection (see Fig. \ref{fig:chainexample}) have been designed and produced at FBK (Trento, Italy). These devices have a pixel-pad layout similar to common hybridisation and module interconnect use cases. Three daisy chain structures were produced and bonded, and they are shown in Fig. \ref{fig:chainperipheral}. They emulate an ASIC with a peripheral pad layout on a 20 mm by 20 mm glass substrate. In total, 536 connection pads, measuring 88 $\mu$m by 88 $\mu$m with a 140 $\mu$m pixel pitch, provide the necessary connections for the device. The ACA used for these tests were ACF1 and ACF2 and one ACP1 with 4 $\mu$m diameter silver coated conductive particles embedded in Araldite2011. ACP1 uses the smallest particles available for in-house mixing as shown in Table \ref{tab:Materials}. ACF1 and ACF2 use particles of 3 $\mu$m diameter and a particle density of 71k/mm${^2}$ and 60k/mm${^2}$ . ACP1 at a volume density of 5\% has an equivalent particle density to the ACF of 20-25k/mm${^2}$ (accounting for the different thicknesses of the ACF). Under the assumption of a parallel connection of particles and neglecting possible pad to pad connections we would expect an increase of electrical resistance. All pads that are covered by the adhesive and have intact readout lines for measuring the electrical resistance show a connection. The electrical resistance R is most strongly impacted by the wire length to the readout/measurement pads near the bottom of the device. In the graph the resistances are shown as a function of wire length for each readout connection. The value of electrical resistance per connection chain is calculated to be below 5 $\Omega$, after deducting the line resistivity gained from the linear regression of resistances. A measurement of precise enough electrical resistance with the used measurement setup is not able to verify a difference in resistance due to larger particles being used for the ACP sample. The ACP sample shows a larger statistical spread than the ACF samples, which we expect for the sample with the lowest particle density. In a parallel connection of N particles with the electrical resistance $R_p$ the total electrical resistance R is given by $R = R_p / N$. This non linear behaviour of a parallel connection with respect to the number of connections N causes a larger spread in the ACP1 sample with lower particle density.

\subsection{LGAD hybridisation}

\begin{figure}[htbp]
\centering
\includegraphics[width=.6\textwidth]{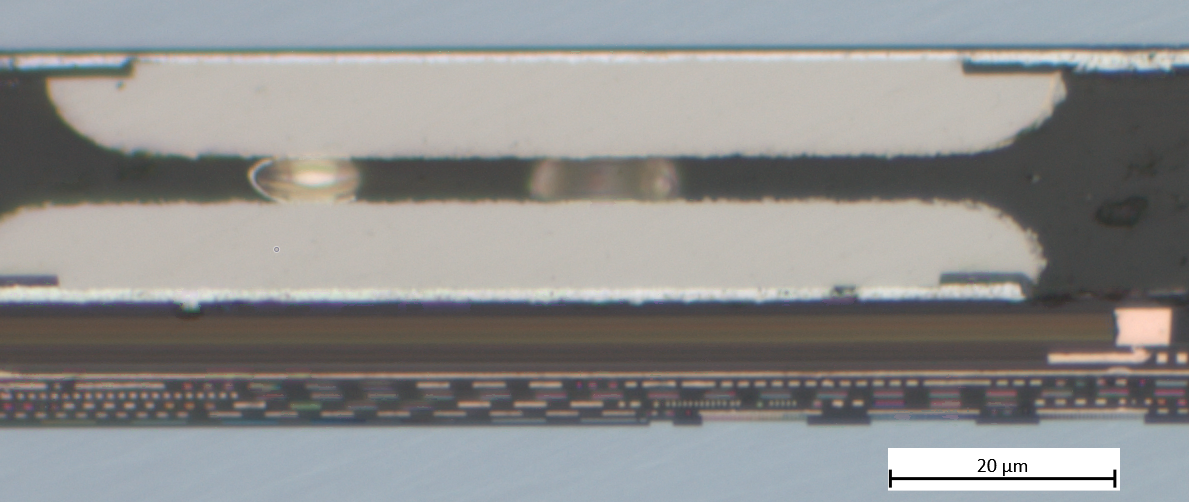}
\caption{Microscopy picture of the cross-section for an ALTIROC2/LGAD device. The connection shown is from row 10 and pad number 7. It shows clearly two particles of 10 $\mu$m diameter being strongly squished between two ENIG coated pads. 
\label{fig:ALTIROCcross}}
\end{figure}

Developing and testing interconnection of large pitch modules has been done as an alternative for solder bump bonding where additional mechanical strength is required. The ALTIROC2 ASIC\cite{ALTIROC2} as well as the Low Gain Avalanche Detector (\textbf{LGAD})\cite{LGAD} sensors produced by USTC (Hefei, China), have a pixel matrix of 15x15 pixels. The pixel pitch is 1.3 mm and the connecting area per pad is around 8000 $\mu$m². The adhesive used for bonding is ACP3 with Ag coated 10 $\mu$m diameter particles. Mechanical samples for cross section microscopy and functional samples have been produced this way. A picture of the cross-section is shown in Fig. \ref{fig:ALTIROCcross}. The distances between pads in the cross-sections of multiple rows spread over the device are measured to be between 0 $\mu$m and 6 $\mu$m. The pads in the centre show the largest distance between pads while the pads near the edge show a closer spacing with some having direct connections between pads. For all the measured pads a connection is expected as the conductive particle are larger in size that the gap they need to bridge. The tested functional assembly shows a connection yield of above 98\%.

\subsection{SPHIRD hybridisation}

\begin{figure}[htbp]
\centering
\includegraphics[width=.95\textwidth]{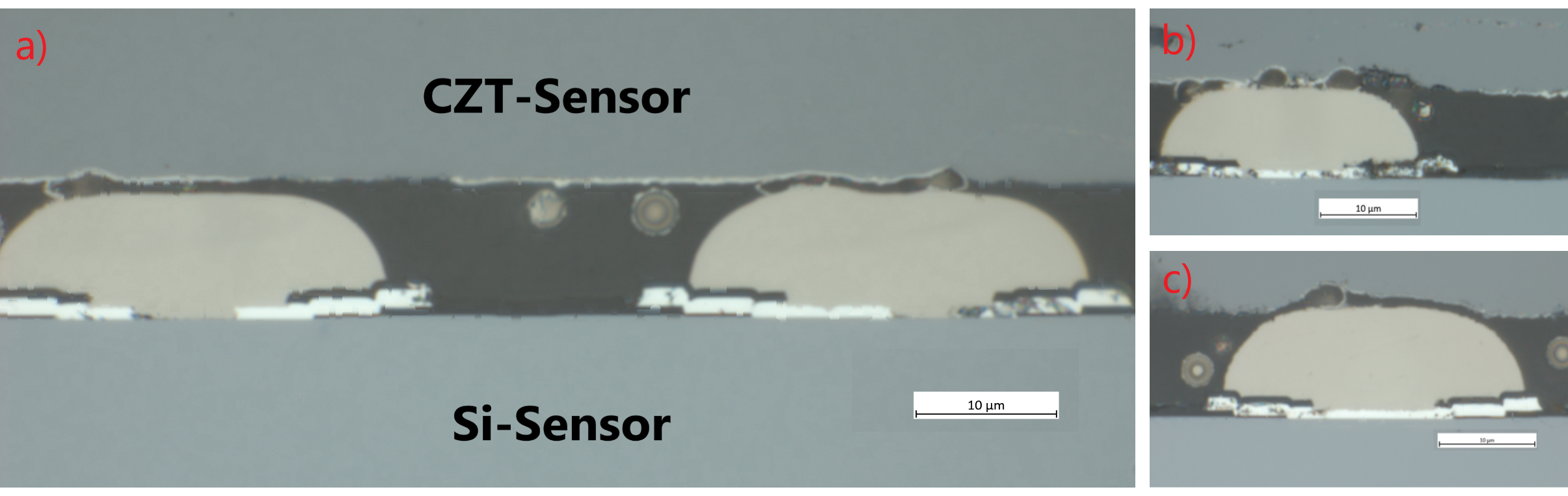}
\caption{Microscope pictures of a cross-sections performed on mechanical assemblies of silicon and a CZT SPHIRD sensor, where only the silicon sensors have been ENIG plated. Particles can be seen in the small gap between CZT-sensor and ENIG pads. a) Shows compressed particles with little deformation on the CZT-sensor side. b) Particles are penetrating the CZT-sensor fully while keeping a spherical shape. c) Particle and ENIG bumb deforming the CZT-sensor substantially.
\label{fig:SPHIRD}}
\end{figure}

Pixel detector developments for the SPHIRD\cite{SPHIRD} project of X-ray detectors at the European Synchrotron Radiation Facility (\textbf{ESRF})(Grenoble, France) use high-Z sensor materials. The hybridisation of sensors made of such materials like CZT has proven to be challenging due to the required single-chip processing as well as temperature and pressure sensitivity of the sensors. The ASIC and sensors used in this study contain a pixel matrix of 32x64 pixels with a pitch of 50 $\mu$m. The ASIC is silicon based while various sensor materials (Si, CdTe, CZT)) are used. First preliminary beam test results for an ACF bonded ASIC to silicon sensor made with ACF2 have been achieved. The results of the test-beam study show that 85\% of the pixels show a response expected for a good connection while about 7\% show a weak response. 7\% show no response. Mechanical sample testing has been performed and analysed via microscopy. The mechanical sample tests use an ENIG plated silicon sensor (as ASIC substitution) and an unplated CZT sensor to test bonding on a sample with plating only on the ASIC side. The cross-section pictures in Fig. \ref{fig:SPHIRD} show connections between the pads either via particles or direct pad-to-pad connections. For Fig. \ref{fig:SPHIRD} b) and Fig. \ref{fig:SPHIRD} c) the bonding pressure of 80 MPa shows a strong plastic deformation of the CZT leading to possible failures caused by disruptions in the lattice structure. The high pressure is amplified for Fig. \ref{fig:SPHIRD} c) as the assembly shows a pressure spike due to nonplanarity in the bonding. Origins of a nonplanarity failure can range from production, bonding process, epoxy curing to humidity absorption\cite{Frisk}. For process self-planarization\cite{planar} as well reducing epoxy spill we attach polyimide tape on the backside of our substrates for the flip-chip bonding.

\subsection{Timepix3 hybridisation}

Bonding of Timepix3\cite{Timepix3} ASICs to planar silicon sensors of 300 $\mu$m thickness is performed with ACF. The Timepix3 ASIC has a 256x256 pixel matrix with one additional row of guard ring pads. The pixel pitch is 55 $\mu$m and the pad diameter is 12 $\mu$m.  For devices like the Timepix3 the distance from the center of the pixel matrix to the edge of the chip is too large too efficiently squeeze the epoxy out of the matrix. Achieving a sufficient plating height needed to displace the excessive adhesive in gaps between the ENIG pads is challenging.
ASIC-side plating of 7-8 $\mu$m  height above the passivation has been achieved with no defects.
The bonding has been done with ACF1 and an extended pre bonding step. For this the temperature was kept at 80°C for 600 seconds while the maximum bonding force of 100 kgf was applied, to give the adhesive time to squeeze out of the connection. This was done at reduced temperatures to ensure that the epoxy does not cure prematurely. Afterwards the temperature is raised to 150°C for 5s bonding and full curing.

\begin{figure}[htbp]
\centering
\includegraphics[width=.4\textwidth]{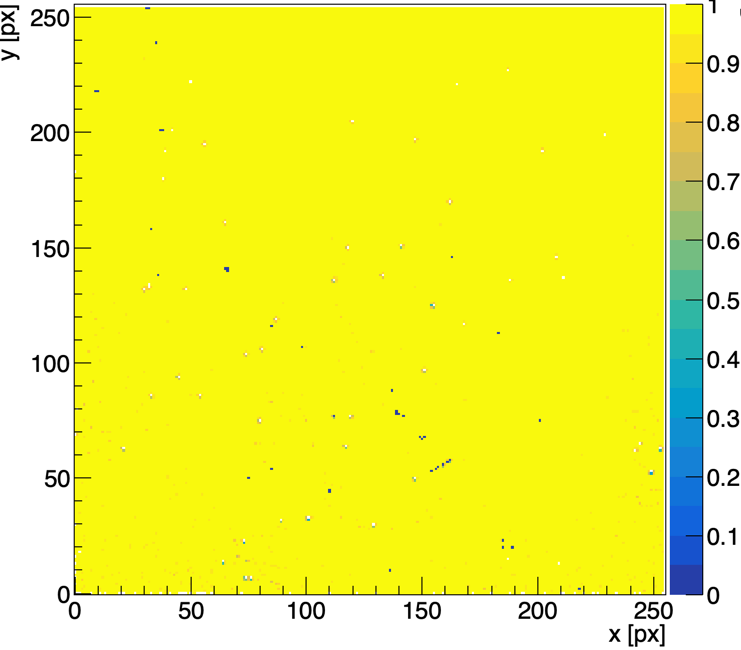}
\qquad
\includegraphics[width=.4\textwidth]{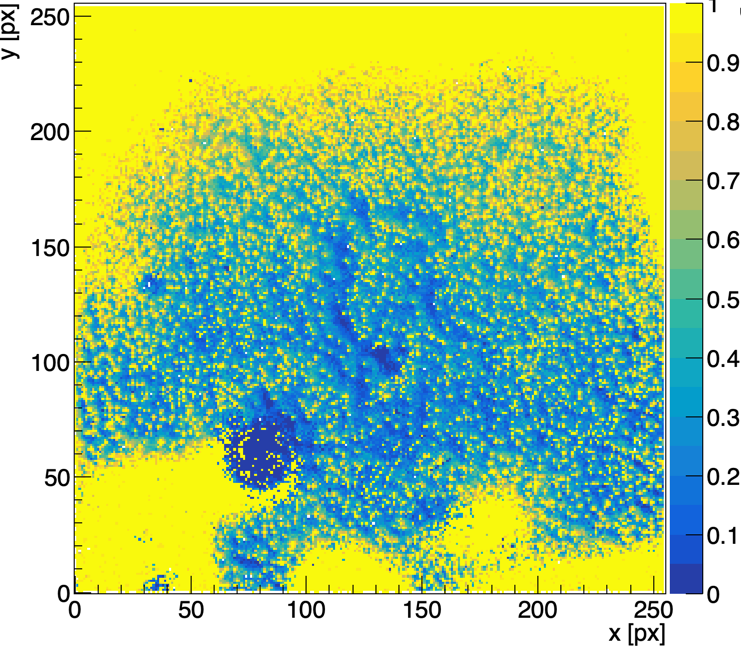}
\caption{Pixel efficiency maps of a Timepix3 assembly, obtained from a test-beam measurement with minimum ionizing particles and a reference beam telescope. The left map shows the detection efficiency for a detection threshold of 700 e-, while the right map shows the detection efficiency for a detection threshold of approximately 4000 e-. 
\label{fig:Timepix3}}
\end{figure}

The produced device was analysed in a test beam measurement. Fig. \ref{fig:Timepix3} shows that moderate increase in detection threshold leads to reduced efficiency in the centre regions of the device, as expected from weakly coupled connections. This is assumed to be caused by too large quantities of adhesive in the centre, which could not be squeezed out of the matrix. The device shows shows an efficiency of 99.54\% for a detection threshold of 700 e- and 33.08\% for a detection threshold of 4000 e-.

\section{Conclusions}
\label{sec:Conclusion}

Electrical measurements on dedicated chain devices confirmed that a low-ohmic connection with high interconnect yield can be achieved for large pitch and large pad areas for both anisotropic conductive films and anisotropic conductive pastes. The results presented for various ASICs and sensors, show promising anisotropic conductive adhesive applications in a wide range of use cases. For large pitch hybridisation it offers an alternative to bump bonding with enhanced mechanical stability provided by the adhesives good interconnect yield. 
The processes under development have proven beneficial for the R\&D efforts of the participating detector projects, due to the fast turnaround and low cost.
The results of the SPHIRD project show a not yet sufficient yield for such small-scale and fine pitch devices. Anisotropic conductive adhesive shows benefits in the bonding of the cadmium zinc telluride (CZT) sensors, where standard solder bump bonding is not useable. The mechanical pressure applied and type of anisotropic conductive adhesive used for the bonding of such sensors still needs adjustments for a reliable interconnection. 
Simulation results of particle distributions show that a higher particle density is required for a sufficient connection yield. For larger matrices such as Timepix3 the effects relating to adhesive quantity and maximum bonding force of the equipment are to be considered. Currently plating heights sufficient and uniform required to displace the amount of adhesive of the films in use are achieved.

A higher ENIG plating and/or thinner anisotropic conductive film or a reduced coverage is needed to a achieve a better connection yield for fine pitch assemblies. More daisy chain test structures are being produced for further bonding tests as well as thermal and mechanical testing in a dedicated reliability study.
Tests with larger particles for anisotropic conductive pastes are planned for large pitch applications.





\acknowledgments

This project has received funding from the European Union’s Horizon 2020 Research and Innovation 
programme under GA no 101004761.



\bibliographystyle{JHEP}


\end{document}